\documentclass[aps,prb,twocolumn,noeprint]{revtex4-1}

\usepackage[bookmarks=false,hyperindex,breaklinks]{hyperref}
\hypersetup{colorlinks=true, citecolor=blue, urlcolor=blue, linkcolor=blue}
\usepackage{graphicx}
\usepackage{epsfig}
\usepackage{amsmath}
\usepackage{graphics}
\usepackage{epsfig}
\usepackage{amsmath}
\usepackage{amsfonts}
\usepackage{amssymb}
\usepackage{xcolor,cancel}
\usepackage{bm}
\usepackage{float}

\usepackage{subcaption}
\PassOptionsToPackage{linktocpage}{hyperref}

\newcommand{\braket}[1]{\ensuremath{\langle{#1}\rangle}}

\begin{document}


\title{Full counting statistics for electron transport in periodically driven quantum dots}

\author{Thomas D. Honeychurch} 
\author{Daniel S. Kosov}
\affiliation{College of Science and Engineering, James Cook University, Townsville, QLD, 4811, Australia}

\begin{abstract}
Time-dependent driving influences the quantum and thermodynamic fluctuations of a system, changing the familiar physical picture of electronic noise which is an important source of information about the microscopic mechanism of quantum transport. Giving access to all cumulants of the current, the full counting statistics (FCS)  is the powerful  theoretical method to study fluctuations in nonequilibrium quantum systems. In this paper, we propose the application of FCS to consider periodic driven junctions. The combination of  Floquet theory for time dynamics and nonequilibrium counting-field Green's functions enables the practical formulation of  FCS for the system. The counting-field Green's functions are used to compute the moment generating function, allowing for the calculation of the time-averaged cumulants of the electronic current. The theory is illustrated using different transport scenarios in model systems.

\end{abstract}
\maketitle

\section{Introduction} 


Time-dependent phenomena play an important part in the investigation and application of nanoscale electronics. The dynamical response of a junction to the modulation of a voltage or to the irradiation by a light source offers intriguing means of probing and controlling the system's dynamics \cite{Platero2004}.
Applications include optically irradiated nanoscale junctions \cite{Meyer2007,Chauvin2006}, electron pumping \cite{Yadalam2016,Croy2016,Haughian2017}, pump-probe spectroscopy \cite{Selzer2013,Ochoa2015}, and AC current rectification \cite{Tu2006,Trasobares2016}.
The exploration of time-dependent phenomena has a long history \cite{Platero2004,Tien1963}, which mostly centers on periodic drivings, typically induced with microwave radiation. For periodic driving, the explanation of photon-assisted transport (PAT) is often invoked: electrons, moving between regions under periodic driving, are observed to undergo inelastic tunneling events, with the absorption and emission of quanta of energy given by the driving frequency \cite{Platero2004}. PAT has been realized within many systems \cite{Dayem1962,Meyer2007,Chauvin2006}, and has undergone extensive theoretical investigation \cite{Platero2004}.


Many approaches have been developed to deal with driving within transport settings. Approaches include Floquet  and  scattering theories \cite{Pedersen1998,Grifoni1998,hanggi2003,Kohler2005,blanter2007,moskalets2002,moskalets2004,moskalets-book,moscalets2014}, quantum master equations \cite{PhysRevB.96.085425,Peskin2017}, nonequilibrium Green's functions (NEGF) based approaches \cite{Arrachea2005,Arrachea2006,Ke2010,Tuovinen2014,Wang1999,Sun1998,Jauho1994,Datta1992,Stafford1996,Brandes1997,Stefanucci2004,Pastawski1992,Chen1991,Maciejko2006,Zhu2005,Ridley2015,Ridley2017,nazarov2007,vonOppen2016,Kienle2010,Sun1997,Honeychurch2020}, and the evolution
operator method \cite{Kwapinski2002,Taranko2004,Kwapinski2005}.


The discrete nature of the charge carrier results in irrepressible current fluctuations. These current fluctuations can contain information about the nature of the junction. For systems with explicit time dependence, understanding current fluctuations is crucial for metrological and performance reasons \cite{Pekola2013,Peskin2017}. Noise has been investigated with Floquet theories \cite{hanggi2003,Suzuki2015}, NEGFs \cite{Feng2008,Ma2004,Chen2008,Sun2000,Zhao2007,Zhao2011,Wu2010NEGF}, and Floquet master-equation approaches \cite{Wu2010}. 


Full counting statistics (FCS) allows the calculation of higher cumulants of the current in a rather concise and systematic manner.
The theory of FCS was first introduced by Levitov and Lesovik  \cite{Levitov1993,Levitov1996} in a scattering theory framework via the explicit inclusion of a measuring device in the theoretical setup, which has since been extended to a general quantum mechanical variable\cite{Esposito2009} and applied to various transport scenarios \cite{nazarov_blanter_2009}. For periodically driven systems, FCS has been coupled within scattering theory \cite{Ivanov1997,Zhang2009,Ivanov2010} and master-equation approaches \cite{Benito2016,Croy2016,Zheng2013} with various investigated periodically driven systems.

NEGFs have proved indispensable  in calculating FCS in quantum conductors with   electron-phonon \cite{Fransson2010,Park2011,Ueda2017,Avriller2009,Schmidt2009} and electron-electron \cite{Gogolin2006,Ridley2018} interactions, and it has been applied to study fluctuations in the transient regime \cite{Tang2014,Yu2016,Ridley2018}.
Recently, Yadalam and Harbola used  FCS NEGFs to study  charge pumping through a driven quantum dot \cite{Croy2016,Yadalam2016}.

The NEGF extension  of FCS to quantum conductors  with arbitrary time-periodic driving of the leads remains the challenging theoretical problem and has not been accomplished yet—this  is the main goal of our paper.
The need for a counting field  complicates derivations within a time-dependent setting, as the equations of motion for the Green's functions become unwieldy integro-differential equations within the Keldysh-Schwinger space. 
In this paper, using Floquet Green's functions\cite{Cabra2020,Brandes1997,Tsuji2008,Haughian2017}, we investigate full counting statistics for modeling systems with  periodic driving of the leads. The periodicity within the system's dynamics allows for the Green's functions of the system to be cast as Fourier series. This allows for the recasting of the equations of motion in terms of matrix equations of infinite dimension, which are truncated at a particular point. Truncating the matrices amounts to truncating the Fourier series describing the Green's functions of the system. This approximation is well behaved and is applicable to situations of both strong and weak coupling to the leads, relative to the chosen driving frequency.


The method was applied to a single resonant level and a T-shaped double-level system. In both cases, the left lead underwent the sinusoidal driving of its energies. This results in photon-assisted transport being observed within the current cumulants, most notably resulting in additional noise in particular cases. For the T-shaped junction, Fano-resonance effects were found to manifest in the peaks generated by the driving. This was explained by considering the bonding and anti-bonding levels of the central molecule.


This paper is organized as follows. Section II describes the use of Floquet Green's functions to solve the equations of motion in an FCS context. In Sec. III, the method is applied to the given systems. In Sec. IV, the results of the paper are summarized. Natural units for quantum transport are used throughout the paper, with $\hbar$, $e$, and $k_B$ set to unity.  

\section{Theory} 

\subsection{General considerations} 

Within the investigation, the Hamiltonian for a central region (i.e., the molecule) connected to two macroscopic leads is given by
\begin{equation}
H(t)=H_C(t) + H_L(t) +H_R(t) +H_{CL}(t) + H_{CR}(t),
\label{hamiltonian}
\end{equation}
where $H_C$ is the Hamiltonian of the molecule, $H_L$  and  $H_R$ are the Hamiltonians of the left and  right leads, and $H_{CL/R}$  is the interaction between the central region and the left/right lead, respectively. 

We consider the electrons within the central region to be noninteracting:
\begin{equation}
H_C = \sum_{ij} h_{ij}(t)d^\dag_i d_{j},
\end{equation}
where $d^\dag_i$ ($d_i$) creation (annihilation) operators are for an electron in the single-particle state $i$. 
 
The left and right leads  are modeled as macroscopic reservoirs of noninteracting electrons,
\begin{equation}
\label{leads}
H_L +H_R =  \sum_{k,\alpha=L,R}  \epsilon_{k\alpha}(t)  c^{\dagger}_{k\alpha} c_{k\alpha},
\end{equation}
where $c^{\dagger}_{k\alpha}$($c_{k\alpha}$) creates (annihilates) an electron in the single-particle state $k$ of either the left ($\alpha=L$) or  the right ($\alpha=R$) lead. The coupling between the central region and left and right leads         {is} given by the tunneling interaction
\begin{equation}
\label{coupling}
H_{CL}+H_{CR}=  \sum_{i, k, \alpha=L,R } \left[t_{k\alpha i}(t) c^{\dagger}_{k \alpha} d_i +\mbox{H.c.}  \right],
\end{equation}
where $t_{k\alpha i}(t)$ is the time-dependent tunneling amplitude between  leads and central single-particle states. 

Utilizing full counting statistics, we follow a common procedure \cite{Gogolin2006,Yadalam2016}. The generating function,
\begin{equation} \label{eq:generating_function1}
\chi \left(\lambda_L, t_c,t_0\right) = \braket{e^{-i\lambda_L N_L \left(t_c\right)}e^{i\lambda_L N_L\left(t_0\right)}},
\end{equation}		
allows for the calculation of the charge that leaves the left lead between times $t_0$ and $t_c$. Here, $N_L \left(t\right)$ is the occupation of the left lead within the Heisenberg picture. The cumulants of the above are given by
\begin{equation} \label{eq:cumulant_generator}
\braket{\delta^n q(t_c,t_0)} = \left. \left(-i\right)^n \frac{\partial^n}{\partial \lambda^n_L} \ln \chi \left(\lambda_L,t_c,t_0\right) \right|_{\lambda_L = 0}.
\end{equation}
By taking the first derivative with respect to the counting field, this result can be recast:
\begin{multline} \label{eq:genF1}
-i \frac{\partial}{\partial \lambda_L} \ln \left[\chi \left(\lambda_L, t_c, t_0 \right)\right] 
\\
= 
\int_{t_0}^{t_c} dt \int_{t_0}^{t_c} dt' \; \operatorname{Tr} \left[ \bm \Sigma^L_{<} (t,t')  \bm G_{>}\left(t',t\right) \right. \\ \left.  - \bm G_{<}\left(t,t' \right) \bm \Sigma^{L}_{>} (t',t) \right],
\end{multline}
where the Hamiltonian is modified by the appropriate counting field 
\begin{equation}
H^{\lambda_L}_{CL} (t) = \sum_{i,k} t_{kL,i}(t) e^{-\frac{i}{2} \lambda_L(\tau)} c^{\dagger}_{k \alpha} d_i + t^*_{kL,i} (t) e^{\frac{i}{2} \lambda_L(\tau)} d^\dagger_i  c_{k \alpha},
\end{equation}
and where $\lambda (\tau) = \pm \lambda_L$ on the forward and backward branches of the contour, respectively. Here, the Green's functions are projections of the contour Green's function with the addition of the counting field:
\begin{equation}
G_{i,j}(\tau,\tau') = -i \frac
{ \operatorname{Tr} \left[ \hat\rho_0 \operatorname{T_c} \left( e^{-i\int_c V_h^{\lambda (\bar{\tau})}(\bar{\tau}) d\bar{\tau}} a_i (\tau) a^\dagger_j (\tau') \right)\right]}
{ \operatorname{Tr} \left[ \hat\rho_0 \operatorname{T_c} \left(e^{-i\int_c V_h^{\lambda (\bar{\tau})}(\bar{\tau}) d\bar{\tau}} \right)\right]},
\end{equation}
where $V^{\lambda (t)}_h$ is the coupling between the leads and the central region within the interaction picture and the additional counting field [i.e. $V^{\lambda (t)}_h = H^{\lambda (t)}_{CL} (t) + H_{CR} (t)$].
The counting field modifies the lesser and greater lead self-energies such that $\bm \Sigma^<_{L} \rightarrow \bm \Sigma^<_{L} e^{i\lambda_L}$ and $\bm \Sigma^>_{L} \rightarrow \bm \Sigma^>_{L} e^{-i\lambda_L}$, while the other self-energy terms are unchanged.

Moving to real time, the contour Green's function is expressed in Schwinger-Keldysh space, 
\begin{equation}
\hat{\bm{G}}(t,t') =
\begin{pmatrix}
\bm{G}^T (t,t') & \bm{G}^< (t,t') \\
\bm{G}^> (t,t') & \bm{G}^{\widetilde{T}} (t,t')
\end{pmatrix},
\end{equation}
with the Kadanoff-Baym equations following the standard definition: 
\begin{multline} \label{eq:Dyson1}
\left(i\frac{\partial}{\partial t} - \bm h(t)\right) \check{\bm G} \left(t,t'\right) - \int^{t_c}_{t_0} dt_1 \check{\bm \Sigma} \left(t,t_1\right) \check{\bm G}\left(t_1,t'\right) \\  = \bm \delta \left(t-t'\right), 
\end{multline}
with $\check{A}_{2,j} \left(t,t'\right) = - \hat A_{2,j} \left(t,t'\right)$. \cite{rammer_2007}

\subsection{Floquet Green's functions}

The system's periodicity within time allows for objects of interest to be cast as Fourier series \cite{Brandes1997,Tsuji2008,Haughian2017}. For a given two-time object, the Fourier coefficients are calculated as
\begin{multline} \label{eq:A in Floquet}
A(\omega, m) = 
\\
 \frac{1}{P}\int_{0}^{P} dT e^{-i\Omega m T} \int_{-\infty}^{\infty} d\tau e^{i\omega\tau} A\left(T,\tau\right),
\end{multline}
where $T= \frac{t+t'}{2}$ and $\tau = t-t'$. Given the above, one can make judicious choices about how to express the Fourier coefficients of the Kadanoff-Baym equations, leading to the problem being cast as matrix equations of infinite dimension.

Here, we outline the procedure of transforming the convolutions into the multiplication of matrices \cite{Brandes1997}. The simpler elements of the Kadanoff-Baym equations follow easily. Considering Eq. (\ref{eq:A in Floquet}), we wish to find the Fourier coefficients of the following object:
\begin{equation} \label{eq:convolution}
C(t,t') = \int dt_1 A(t,t_1) B(t_1,t'),
\end{equation}
given the periodicity of terms $A(t,t')$ and $B(t,t')$ around the central time, i.e., $T = \frac{t+t'}{2}$.
One can begin the transformation of Eq. (\ref{eq:convolution}) by first taking the Wigner transformation:
\begin{equation}
\begin{split}
C(T,\omega) = \int_{-\infty}^{\infty} d\tau e^{i\omega\tau} C(t,t')
\\= 
e^{-\frac{i}{2}\left(\partial_{T}^{A}\partial_{\omega}^{B}-\partial_{\omega}^{A}\partial_{T}^{B}\right)} A \left(T,\omega\right) B \left(T,\omega\right).
\end{split} 
\end{equation}
The objects within Wigner space can be expressed as Fourier series, leading to the simple evaluation of the time derivative:
\begin{multline}
C(\omega,m) 
\\= \sum^{\infty}_{n = 0} e^{\frac{\Omega}{2}\left[n\partial_{\omega}^{B}-\partial_{\omega}^{A}(m-n)\right]} A(\omega,n)  B (\omega,m-n)
\\= \sum_{n = -\infty}^{\infty}  A\left(\omega + \frac{\Omega}{2}(n-m),n\right)B\left(\omega + \frac{\Omega}{2}n ,m-n\right).
\end{multline}
Introducing $\omega \rightarrow \omega + l \Omega /2$, where $l$ is an integer, and making use of the notation $\check{G} (\omega + \frac{\Omega}{2}l,m) \rightarrow \check{G}_{l,m}$, the above becomes
\begin{equation}
\begin{split}
C_{l,m} = \sum_{n = -\infty}^{\infty} A_{l+n-m,n} B_{l+n,m-n}
\\ = \sum_{n' = -\infty}^{\infty} A_{l+n',n'+m} B_{l+m+n',-n'},
\end{split}
\end{equation}
where the last equality is completed with $n' = n - m$. We now complete an index transformation, such that $A_{i,j} \rightarrow \mathcal{A}_{r,s}$, where $r = (i-j)/2$ and $s = (i+j)/2$:
\begin{equation} \label{eq:A4}
\mathcal{C}_{r,s} = \sum_{n' = -\infty}^{\infty} \mathcal{A}_{r,n'+s} \mathcal{B}_{n'+s,s} = \sum_{n = -\infty}^{\infty} \mathcal{A}_{r,n} \mathcal{B}_{n,s},
\end{equation}
where the last equality has a shift in the infinite summation. With the assumption of periodicity, the above convolution has been brought to the form of a matrix equation. These matrices (the Floquet matrices), with indices running from negative infinity to infinity, take the following form in terms of the original Fourier components:
\begin{widetext}
	\begin{equation}
	\bm{\mathcal{A}} = 		\begin{pmatrix}
	\hdots & \hdots &  \hdots & \hdots & \hdots  \\
	\hdots & A(\omega-\Omega,0) & A(\omega-\frac{\Omega}{2},1) & A(\omega,2) & \hdots \\
	\hdots & A(\omega-\frac{\Omega}{2},-1) & A(\omega,0) & A(\omega+\frac{\Omega}{2},1) & \hdots \\
	\hdots & A(\omega,-2) & A_{i,j}(\omega+\frac{\Omega}{2},-1) & A(\omega + \Omega,0) & \hdots \\
	\hdots & \hdots &  \hdots & \hdots & \hdots  \\
	\end{pmatrix}.
	\end{equation}
\end{widetext}

In the context of the problem, the elements of the Floquet matrices are the Fourier coefficients of the Green's function or the lead self-energies in Keldysh-Schwinger space, themselves matrices of $2n$ dimensions. Via a unitary transformation, the elements within the Floquet matrices can be rearranged. As a consequence, they can be brought to look like the matrices one would expect in the static case, but with the innermost elements being infinite matrices populated by Fourier coefficients, as opposed to a scalar within the static case.

Generally, to solve for the inverses of the Floquet matrices, one must truncate the matrices at a large enough dimension to have a negligible effect on the calculations. Removing all time dependence, the matrices reduce to scalars, agreeing with the static case results.

We also have to consider the following limiting cases of Eq. (\ref{eq:convolution}):
\begin{equation}\label{eq:convolution2}
C(t,t') = A(t) B(t,t'),
\end{equation}
which follows from $A(t,t') \rightarrow A(t)\delta\left(t-t'\right)$. Therefore, given the application of Eq. (\ref{eq:A in Floquet}), the above has the same form as in Eq. (\ref{eq:A4}), with the Floquet matrix of $\bm{\mathcal{A}}$ populated with the Fourier coefficient of $A(t)$. 

Solving the Kadanoff-Baym equations by invoking a Floquet approach restricts the method to considering time-averaged statistics: the points of measurement are moved to the infinities to ensure the periodicity of the system (i.e., $t_0 \rightarrow -\infty$ and $t_c \rightarrow \infty$). Following the above considerations, the Kadanoff-Baym equation can be transformed into a matrix equation:
\begin{multline}\label{eq:truncation main}
\left[ 
\begin{pmatrix}
\omega \mathbf{I} & 0 \\
0 & \omega \mathbf{I}
\end{pmatrix} 
+		
\begin{pmatrix}
\Omega  \mathbf{D} & 0 \\
0 & \Omega \mathbf{D}
\end{pmatrix} 
\right.
\\
-
\left.
\begin{pmatrix}
\mathbf{H} & 0 \\
0 & \mathbf{H}
\end{pmatrix}
-
\begin{pmatrix}
\mathbf{\Sigma^T} & \mathbf{\Sigma^<} \\
-\mathbf{\Sigma^>} & -\mathbf{\Sigma^{\widetilde{T}}}
\end{pmatrix}
\right]  \bm{\mathcal{\check{G}}}  = 		
\begin{pmatrix}
\mathbf{I} & 0 \\
0 & \mathbf{I}
\end{pmatrix} ,
\end{multline}
where $(\mathbf{D})_{a,b} = a\delta_{a,b}$, $(\mathbf{H})_{a,b}  = \mathbf{h}(b-a)$, and the elements of the self-energy are given as 
\begin{equation}
(\mathbf{\Sigma})_{a,b} = \mathbf{\Sigma}\left(\omega+ (a+b)\Omega /2, b-a\right),
\end{equation}
for a given projection. See the Appendix for calculations of the self-energies.

One can recast Eq. (\ref{eq:genF1}) in terms of Floquet matrices, giving us
\begin{equation} \label{eq:cumulant_equation}
\begin{split}
C_n =\lim_{t_c,t_0 \rightarrow \pm \infty} \left( \frac{(-i)^n \frac{\partial^n}{\partial \lambda^n_L} \ln \left[\chi \left(\lambda_L, t_c,t_0 \right)\right] }{t_c-t_0} \right)
\\
= 
(-i)^{n-1} \frac{\partial^{n-1}}{\partial \lambda^{n-1}} \int^{\infty}_{-\infty} \frac{d\omega}{2 \pi}
\operatorname{Tr} \left[ X(\omega,0) \right],
\end{split}
\end{equation}
where $n\geq 1$ and $X(\omega,0)$ is the zeroth Fourier component, calculated from the Floquet matrix equation
\begin{equation}
\bm X = \bm \Sigma^L_{<} \bm{\mathcal{G_{>}}} - \bm{\mathcal{G_{<}}} \bm \Sigma^{L}_{>}.
\end{equation}
This result corresponds to calculating the time-averaged cumulants of the current, with $C_1 = I$ and $C_2 = S_{LL}(\omega = 0)$ \cite{Fransson2010,kambly2013}.

\section{Application}

\subsection{Resonant level}

We begin by considering a central region consisting of a single resonant level.  The central region and leads are both taken  to have sinusoidal driving.
The central region Hamiltonian is
\begin{equation}
H_C(t) =  \left[\epsilon_0 +\Delta_{0} \cos \left(\Omega t\right)\right] d^\dag d,
\end{equation}
and the leads' energy levels are
\begin{equation}
\epsilon_{k\alpha} (t) = \epsilon_{k\alpha} + \Delta_{\alpha} \cos \left(\Omega t\right).
\label{tdep}
\end{equation}
For the details of the derivations for the time-dependent lead self-energies, see the Appendix.

Considering the current, one can turn off the counting field and rearrange Eq. (\ref{eq:cumulant_equation}) to find
\begin{multline}\label{eq:exact theory}
I = \sum_{k=-\infty}^{\infty} \int^\infty_{-\infty} \frac{d\omega}{2\pi} \;\; T\left(\omega-k\Omega\right) 
\\
\times\left(f_L\left(\omega\right) J^2_k \left(\frac{\Delta_{0}-\Delta_L}{\Omega}\right) - f_R\left(\omega\right) J^2_k \left(\frac{\Delta_{0}-\Delta_R}{\Omega}\right)\right)
\end{multline}
where
\begin{equation}
T\left(\omega\right) = \frac{\Gamma_L \Gamma_R}{\left(\omega-\epsilon_0\right)^2+ \frac{\Gamma^2}{4}},
\end{equation}
in agreement with the literature \cite{Jauho1994,Tang2014}. We see the established phenomena of photon-assisted transport, where the sinusoidal driving of regions results in an effective splitting of the transmission between pictures which can be interpreted as the absorption or emission of photons of $\hbar\Omega$ \cite{Platero2004}. 

\begin{figure*}[]
	\centering
	\hspace*{-7cm}    
	\begin{subfigure}[]{1in}
		\centering
		\includegraphics[width=3.6\textwidth]{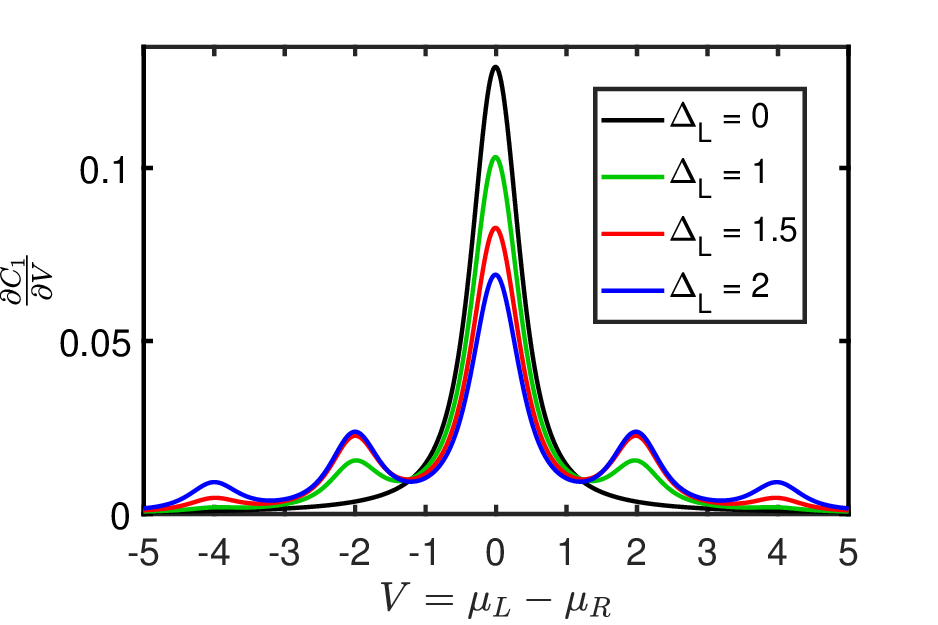}
		\caption{}\label{fig:1a}		
	\end{subfigure}
	\qquad\qquad\qquad\qquad\qquad\qquad\qquad\qquad\qquad\qquad\quad
	\begin{subfigure}[]{1in}
		\centering
		\includegraphics[width=3.6\textwidth]{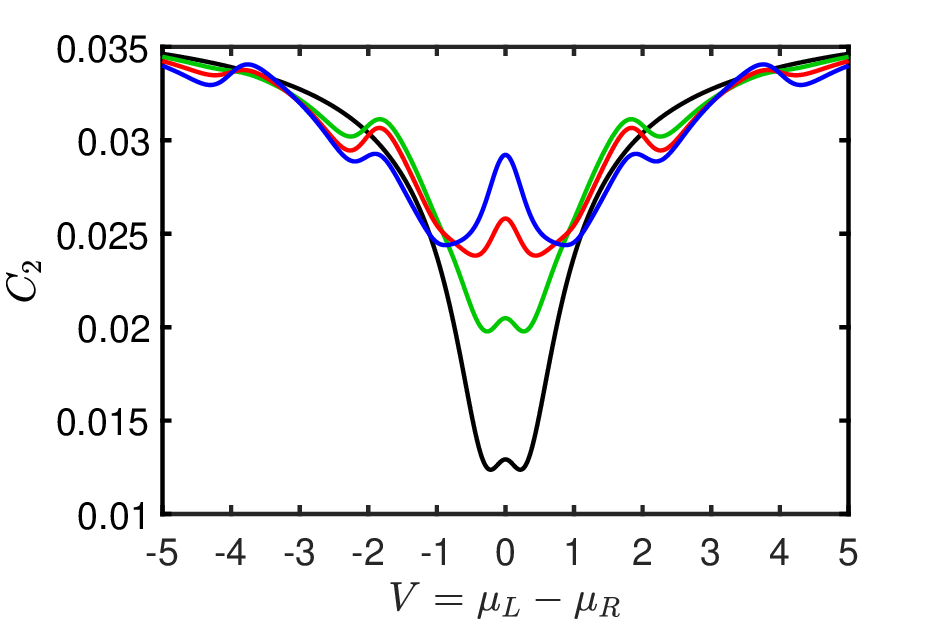}
		\caption{}\label{fig:1b}
	\end{subfigure}
	
	\hspace*{-7cm}   
	\begin{subfigure}[]{1in}
		\centering
		\includegraphics[width=3.6\textwidth]{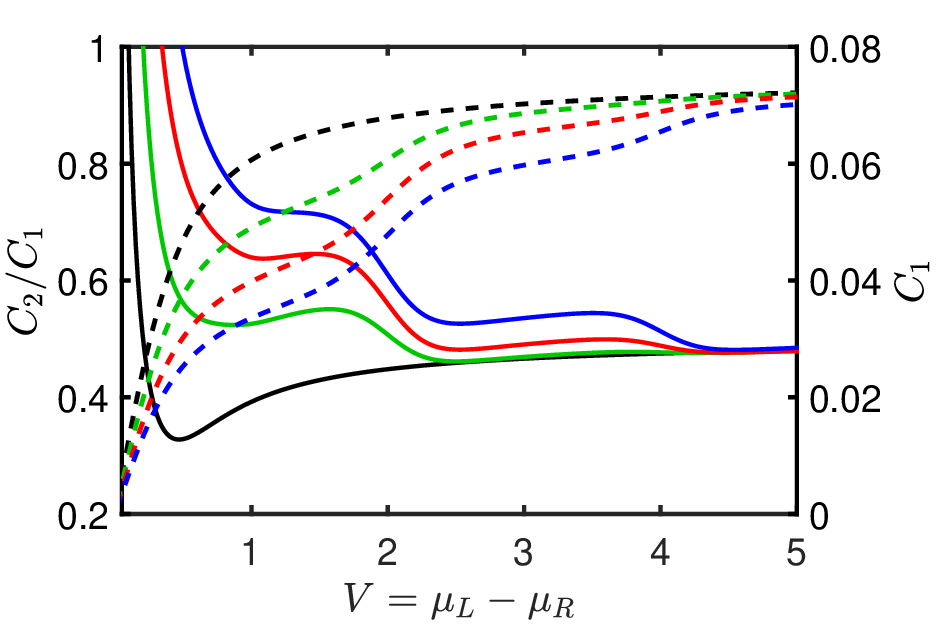}
		\caption{}\label{fig:1c}
	\end{subfigure}
	\qquad\qquad\qquad\qquad\qquad\qquad\qquad\qquad\qquad\qquad\quad
	\begin{subfigure}[]{1in}
		\centering
		\includegraphics[width=3.6\textwidth]{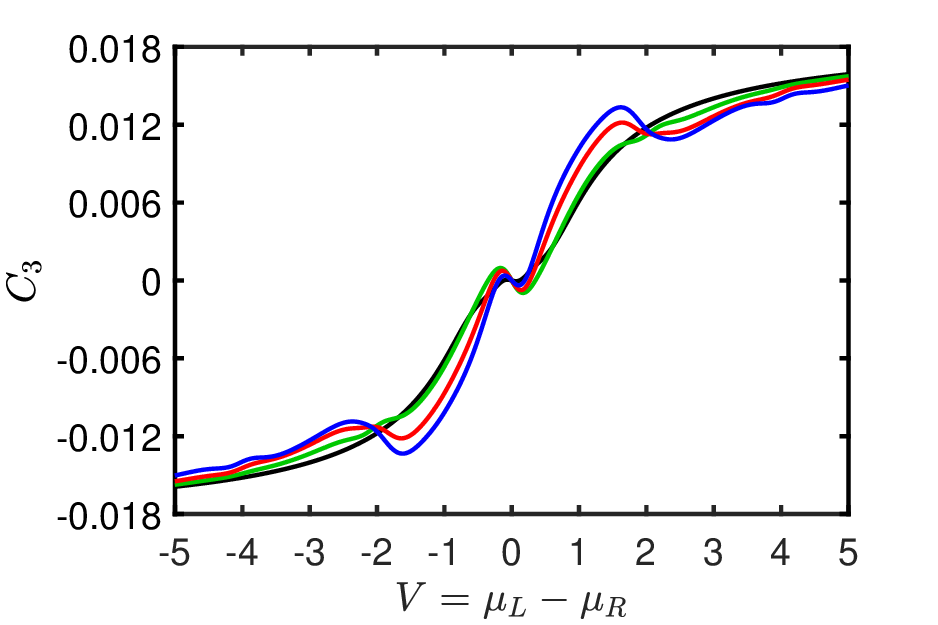}
		\caption{}\label{fig:1d}
	\end{subfigure}
	
	\caption{Cumulants of the current plotted against increasing voltage ($\mu_L=-\mu_R$) for a single level. The left lead's driving is increased, revealing the effects of the photopeaks. Dashed lines shows $C_1$, and solid lines show $C_2/C_1$ in (c). The parameters are $\Gamma_L=\Gamma_R=0.15$, $\varepsilon_1=0$, $T=0.05$, $\Omega=1$, $\Delta_0=0$, and $\Delta_R=0$. }\label{fig:1}
\end{figure*}

\begin{figure}[]
	\centering
	\includegraphics[width=0.5\textwidth]{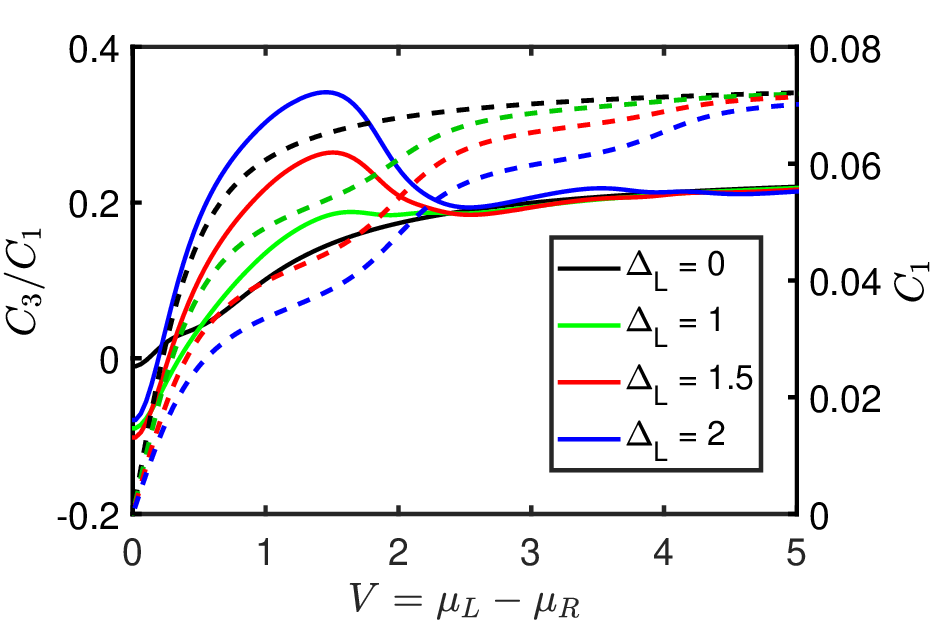}
	\caption{The ratio of cumulants, $C_3/C_1$ plotted against increasing voltage ($\mu_L=-\mu_R$) for a single level. Dashed lines shows $C_1$, and solid lines show $C_3/C_1$. The left lead's driving is increased, revealing the effects of the photopeaks. The parameters are those of Fig. \ref{fig:1}.}\label{fig:1another}
\end{figure}

The results of these calculations are given in Figs. \ref{fig:1} and \ref{fig:1another}. Within all the calculations completed, the matrices were truncated to consider 41 of the Fourier coefficients of each object (i.e., from $n=-20$ to $n=20$). We see that the higher cumulants also display characteristics surrounding the positions of the photopeaks. With the photopeaks entering into the voltage window, the cumulants $C_2$ and $C_3$ decrease in magnitude, suggesting that the moving of the photopeaks into resonance decreases the variability within the average current. Indeed, as the photopeaks move into the resonance, the ratio $C_2/C_1$, an appropriate indicator for the clarity of the current, decreases. This is expected, with increasing voltage, moving the effects of the driving further away, as the level moves further into resonance.

For $C_3$, as the photopeak approaches the voltage window, we see an increase, followed by a decrease, when the photopeak has entered into the voltage window. The above suggests that, around voltages where photopeaks are entering resonance, the current suffers an increased skewness in its distribution. This can be seen with the ratio of $C_3/C_1$, which is larger than in the static case (see Fig. \ref{fig:1another}), within the region before the first photopeak enters resonance.

\subsection{Quantum interference}

Within the central junction, effects due to inference between many levels can give rise to interesting and complicated phenomena\cite{cuevas2010molecular}. In particular, many-level systems, often containing Aharonov-Bohm interference effects, have been studied with time-dependent drivings \cite{Fujisawa1997,Wu2006,Bai2017b,He2018,Oosterkamp1998,Pan2008,Shang2013,Stafford1996,Sun1998,Tang2014,Bao-Chuan2016,Zhao2019,Zhao2014a,Zhao2014b,Zhao2011}. Here, we investigate a simple manifestation of Fano interference due to a secondary offset level coupled to our primary site but uncoupled from the electrodes. The introduction of this secondary site results in interference between the two paths through the system.
The Hamiltonian for the central region is 
\begin{equation}
H_C = \epsilon_1 d_1^\dag d_1 +  \epsilon_0 d_0^\dag d_0 + t (d^\dag_1 d_0 + d_0^\dag d_1),
\end{equation}
and  the central region is connected to the leads through level $\epsilon_0$
\begin{equation}
H_{CL}+H_{CR}=  \sum_{i, k, \alpha=L,R } (t_{k\alpha } c^{\dagger}_{k \alpha} d_0 +\mbox{H.c.} ).
\end{equation}
The leads are  sinusoidally driven  as in the resonant level case considered before Eq. (\ref{tdep}), and the central region Hamiltonian remains static all the time.

Calculations for the current concur with theory [Eq. (\ref{eq:exact theory})], with the transmission undergoing well-known changes due to the introduction of the second level \cite{cuevas2010molecular}:
\begin{equation}
T\left(\omega\right) = \frac{\Gamma_L \Gamma_R}{\left[\omega-\epsilon_0 - t^2 / \left(\omega-\epsilon_1\right)\right]^2+ \frac{\Gamma^2}{4}}.
\end{equation}

\begin{figure*}[t]
	\centering
	\hspace*{-7cm}    
	\begin{subfigure}[]{1in}
		\includegraphics[width=3.6\textwidth]{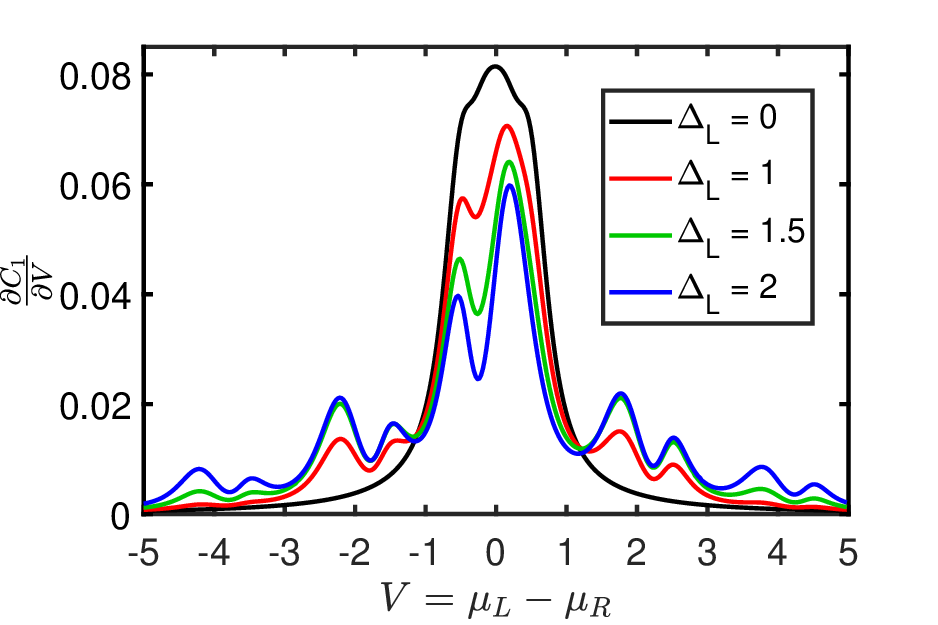}
		\caption{}\label{fig:2a}		
	\end{subfigure}
	\qquad\qquad\qquad\qquad\qquad\qquad\qquad\qquad\qquad\qquad\quad
	\begin{subfigure}[]{1in}
		\includegraphics[width=3.6\textwidth]{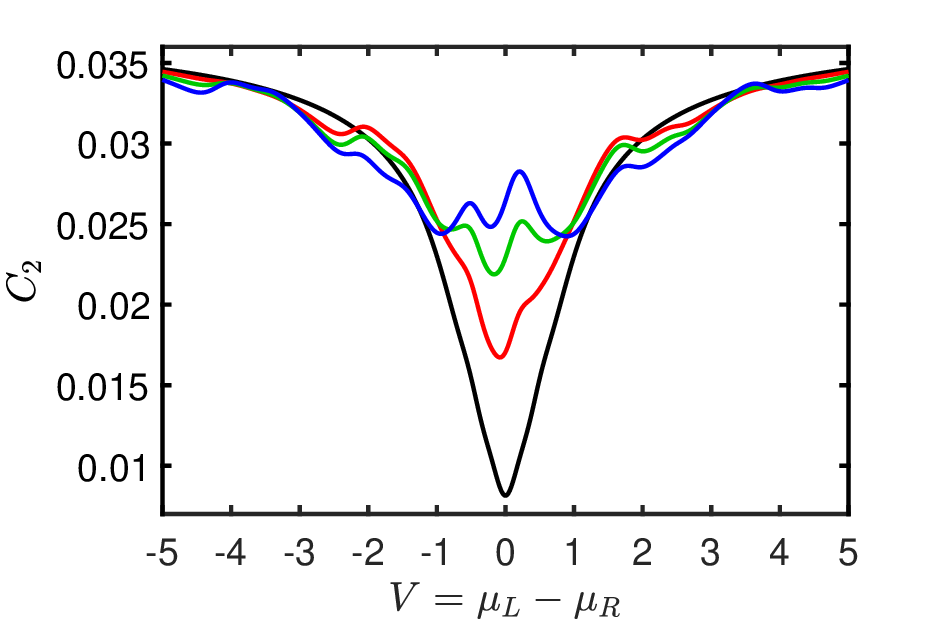}
		\caption{}\label{fig:2b}
	\end{subfigure}
	
	\hspace*{-7cm}   
	\begin{subfigure}[]{1in}
		\includegraphics[width=3.6\textwidth]{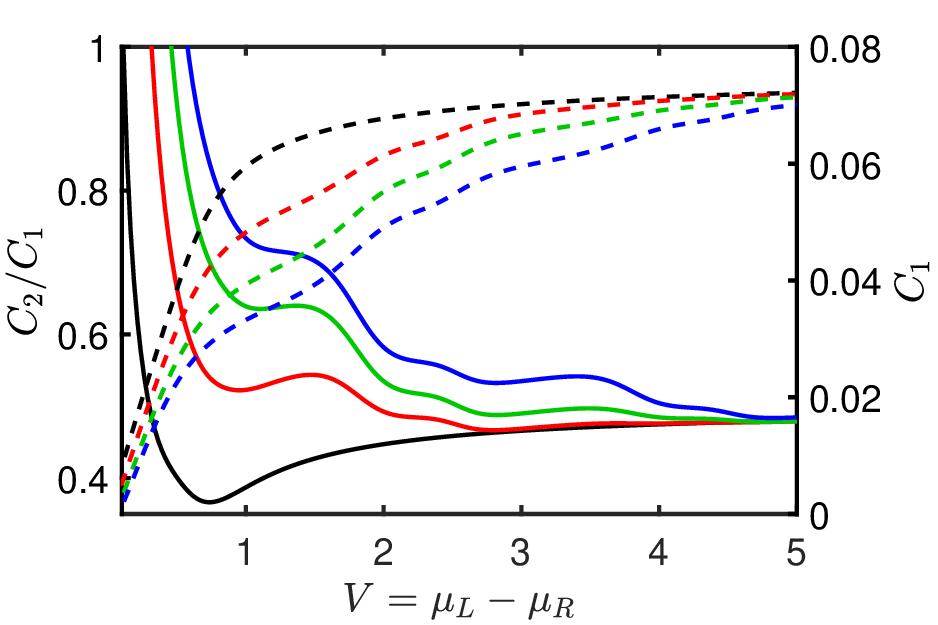}
		\caption{}\label{fig:2c}
	\end{subfigure}
	\qquad\qquad\qquad\qquad\qquad\qquad\qquad\qquad\qquad\qquad\quad
	\begin{subfigure}[]{1in}
		\includegraphics[width=3.6\textwidth]{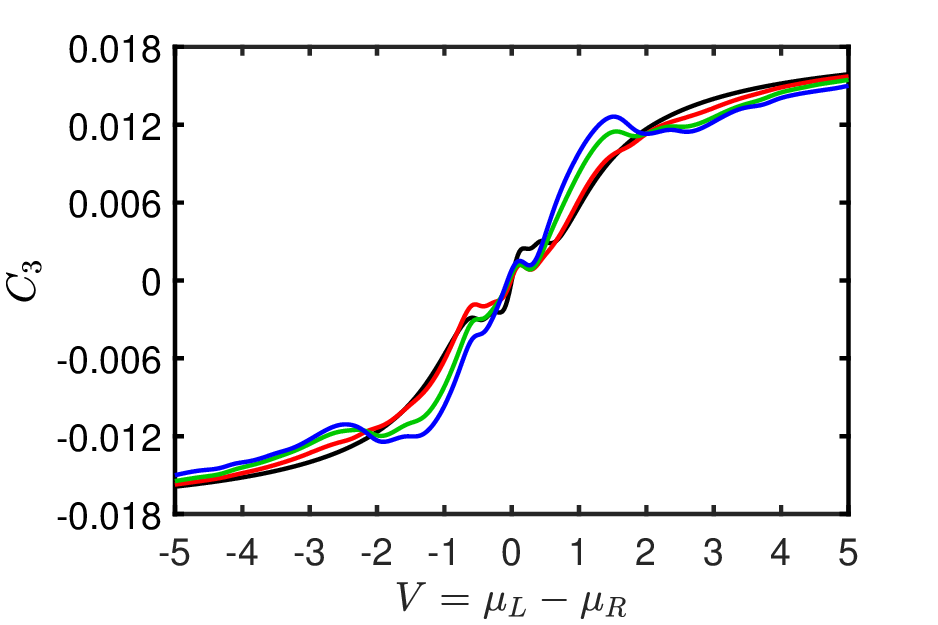}
		\caption{}\label{fig:2d}
	\end{subfigure}
	
	\caption{Cumulants of the current plotted against increasing voltage ($\mu_L=-\mu_R$). The leads are connected to the first level, which is further connected to the isolated second level. Dashed lines shows $C_1$, and solid lines show $C_2/C_1$ in (c). Here, the left lead's driving is increased to reveal the effects of the Fano interference on the photopeaks and higher cumulants. The parameters are $\Gamma_L=\Gamma_R=0.15$, $\varepsilon_0=0$, $\varepsilon_1=0.15$, $t=0.15$, $T=0.05$, $\Omega=1$,  and $\Delta_R=0$.}\label{fig:2}
\end{figure*}

The results of the calculations are given in Fig. \ref{fig:2}. With the introduction of the secondary level, the photon-assisted sidebands in the current acquire pronounced asymmetrical interference patterns from the Fano resonance [see Fig. \ref{fig:2a}], with the asymmetry being governed by the difference between the energies of the levels. Moreover, the asymmetric feature manifests in the central peak differently from within the photopeaks, with a change in the difference in the levels' energies pushing the feature in the opposite direction for the photopeaks compared to the central peak.

To understand the effects of the secondary level, one can diagonalize the central region Hamiltonian, allowing for the investigation of the current in terms of the bonding (-) and antibonding (+) molecular orbitals \cite{Lu2005}, which gives the energies
\begin{equation}\label{eq:molecular energies}
\epsilon_{\pm} = \frac{1}{2} \left(\epsilon_0+\epsilon_1 \pm \sqrt{\left(\epsilon_0-\epsilon_1\right)^2 + 4 t^2}\right)
\end{equation}
and a modified linewidth function:
\begin{equation}\label{eq:modified linewidth}
\widetilde{\bm\Gamma}^{\alpha}_{i,j} = \Gamma^{\alpha} \begin{pmatrix}
\cos^2(\beta) & \cos(\beta)\sin(\beta) \\
\cos(\beta)\sin(\beta)  & \sin^2(\beta)
\end{pmatrix},
\end{equation}
where
\begin{equation}
\beta = \frac{1}{2} \tan^{-1}\left(\frac{2t}{\epsilon_0 - \epsilon_1}\right).
\end{equation}

The diagonalization of the central region allows us to identify the roles that the bonding and antibonding molecular levels play in the cumulants. Furthermore, the effects due to the interaction between these levels is relegated to the off-diagonal terms of the linewidth function [Eq. (\ref{eq:modified linewidth})], with their removal corresponding to a system without interaction between the bonding and antibonding levels.

\begin{figure}[]
	\centering
		\includegraphics[width=0.5\textwidth]{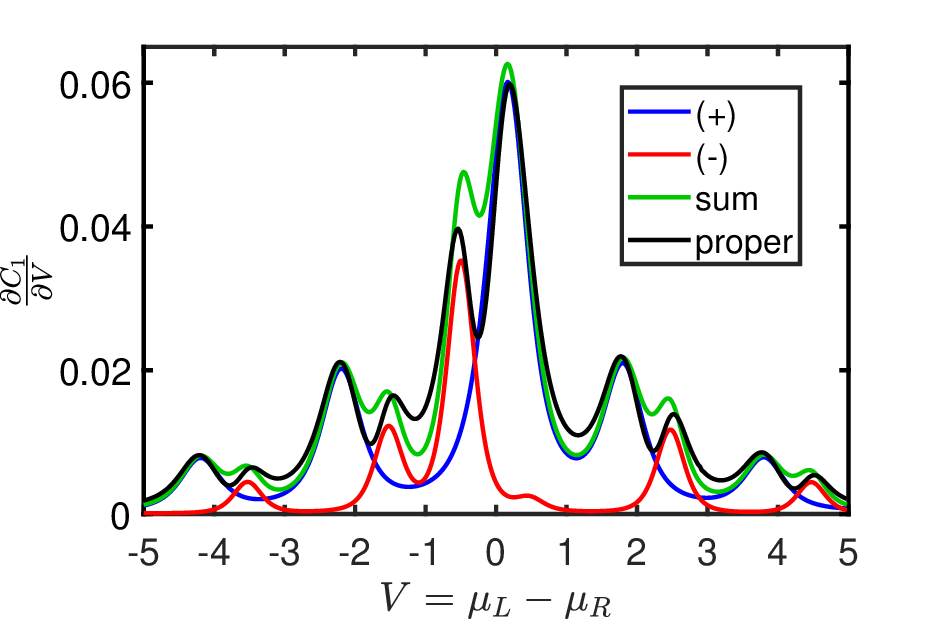}
	\caption{The differential conductance plotted against increasing voltage ($\mu_L=-\mu_R$). The leads are connected to the first level, which is further connected to the isolated second level. Here, the bonding and antibonding levels are plotted, disregarding the off-diagonal terms within the diagonalized linewidth function [Eq. \ref{eq:modified linewidth}], along with their sum and the full result, which does include the off-diagonal terms. The parameters are $\Gamma_L=\Gamma_R=0.15$, $\varepsilon_0=0$, $\varepsilon_1=0.15$, $t=0.15$, $T=0.05$, $\Omega=1$,  and $\Delta_R=0$.}\label{fig:3}
\end{figure}

The effects of the separation into the bonding and antibonding levels can be seen within Fig. \ref{fig:3}. Here, the contributions by both levels are plotted for the case where the off-diagonals within the modified linewidth function [i.e. Eq. (\ref{eq:modified linewidth})] are disregarded. We see that the contributions from the diagonalized levels approximately explain the positions of the peaks within the differential conductance, with the off diagonal terms of the modified linewidth varying the final positions slightly. This analysis also helps to explain the features of the higher cumulants in terms of the bonding and antibonding molecular levels. 

The effects of the asymmetrical feature due to the Fano resonance can be seen to effect the higher cumulants [see Fig. \ref{fig:2b} and \ref{fig:2d}]. While this complicates the features of the cumulants, it was found to not alter the junction's dynamics significantly. This is evident in $F_2 = C_2 / C_1$ [see Fig. \ref{fig:2c}], which suggests no significant changes in the efficiency of the device, with the introduction of a second level resulting in a splitting of the plateauing effect, seen in Fig. \ref{fig:1c}, due to the shared influence of the bonding and antibonding levels. 

\section{Discussion and Conclusion} 

The method investigated allows for the calculation of the time-averaged cumulants of the current for periodically driven molecular junctions. Expressing the Green's functions in terms of a Fourier series allows for the equation of motion to be cast as a matrix equation of infinite dimensions, which, following truncation, can be easily solved. The method was applied to investigate the time-averaged current cumulants for both a single-level system and a T-shaped double-level system. In investigating the higher cumulants the derivatives of the counting field were evaluated with the finite central difference method.

The method is applicable to many levels, and the wideband approximation is not essential. Furthermore, its application to systems considering extra correlations is conceivable. 

It should be noted that the above method calculates the time averages of the cumulants of the current. These calculations should not be confused with the statistics one could calculate when considering the distribution of instantaneous values (for current, noise, etc.) that could be calculated within a period of the driving. While the latter would be more appropriate if one were interested in the instantaneous measurements, the former is more appropriate for devices which would cumulate large transmitted charges before measurement. 

Within the investigation, it was found that the time-dependent driving of the leads not only induces photopeaks within the current but also generates features within the higher cumulants. The positioning of photopeaks relative to the voltage window was found to have a significant effect on the higher cumulants. Photopeaks that sit just outside resonance have the effect of increasing both the second and third cumulants, while photopeaks that sit just within resonance see a relative decrease in both cumulants. Furthermore, it was found that the time-dependent driving broadened and skewed the current distribution. This was observable within $C_2/C_1$ and $C_3/C_1$ [see Figs. \ref{fig:1c} and \ref{fig:1another} respectively].

Even within the reduced picture of considering time averages, it is evident that the effects of time-dependent driving go beyond the current. This suggests that further investigations into the cumulants of important observables will be essential for understanding time-dependent driven systems.

On top of this, methods that can be extended to consider extra correlations (i.e., electron-phonon and electron-electron interactions) will allow for a fuller understanding of time-dependent driven molecular junctions as they are truly realized, which will hopefully lead to further interesting results.

\newpage
\appendix 

\section{Lead self-energies with the AC driving} \label{App:B}

The lead self-energies are given as 
\begin{equation} \label{eq:self energy definition}
\Sigma^{<,>,T,\widetilde{T}}_{\alpha,ij} (t,t') = \sum_{k,k'} t^*_{k\alpha,i}(t)  \; g^{<,>,T,\widetilde{T}}_{k\alpha,k'\alpha}\left(t,t' \right)  t_{k'\alpha,j}(t'),
\end{equation}
where 
\begin{equation}
g_{k\alpha,k' \alpha'}^< (t,t') = i  f_{k\alpha} e^{-i \int_{t'}^{t} dt_1 \varepsilon_{k\alpha} (t_1)} \delta_{k, k'},
\end{equation}
\begin{equation}
g_{k\alpha,k'\alpha}^> (t,t') = -i (1-f_{k\alpha}) e^{-i \int_{t'}^{t} dt_1 \varepsilon_{k\alpha} (t_1)} \delta_{k, k'},
\end{equation}
\begin{equation}
\begin{split}
g^T_{k,k'} (t,t') =  g_{k\alpha,k'\alpha}^> (t,t') \Theta \left(t-t'\right) 
+ g_{k\alpha,k' \alpha}^< (t,t') \Theta \left(t'-t\right),
\end{split}
\end{equation}
\begin{equation}
\begin{split}
g^{\widetilde{T}}_{k,k'} (t,t') =  g_{k\alpha,k'\alpha}^< (t,t')  \Theta  \left(t-t'\right) 
+ g_{k\alpha,k' \alpha}^> (t,t')  \Theta  \left(t'-t\right).
\end{split}
\end{equation}
The Fermi-Dirac occupation is given by the standard definition:
\begin{equation}
f_{k\alpha} = \frac{1}{1 + e^{(\epsilon_{k\alpha}-\mu_\alpha)/T_\alpha}}.
\end{equation}

Within the investigation, we assume sinusoidal driving within the leads,

\begin{equation}
\varepsilon_{k\alpha} (t) = \varepsilon_{k\alpha} + \Delta_\alpha \cos (\Omega_\alpha t),
\end{equation}

and that the couplings to the leads are constant. We can collect the terms together, making use of the definition for the self energy in the static case (denoted with an apostrophe):
\begin{equation}
\begin{split}
\Sigma_{\alpha,ij} (t,t') 
= \Sigma'_{\alpha,ij} (t - t') e^{-i \int_{t'}^{t} dt_1 \Delta_\alpha \cos (\Omega_\alpha t_1)}
\\=  e^{-i\frac{\Delta_\alpha}{\Omega_\alpha} \sin \left(\Omega_\alpha t\right)} \Sigma'_{\alpha,ij} (t - t') e^{i\frac{\Delta_\alpha}{\Omega_\alpha} \sin \left(\Omega_\alpha t'\right)}.
\end{split}
\end{equation} 
We see that the above follows a pattern similar to Eq. (\ref{eq:convolution}) and, following a similar analysis, can be expressed as the matrix multiplication of three Floquet matrices:
\begin{equation}
\bm \Sigma_{\alpha,ij} = \bm{\mathcal{S}} \bm \Sigma'_{\alpha,ij} \bm{\mathcal{S}}^\dagger.
\end{equation}
Here, $\bm{\mathcal{S}}$ is found with the use of the Jacobi-Anger expansion:
\begin{equation}
e^{i z \sin(\theta)} = \sum_{n=-\infty}^{n=\infty} J_n (z)  e^{in\theta},
\end{equation} 
such that $\bm{\mathcal{S}}_{s,r} = J_{s-r} \left(\Delta_\alpha / \Omega_\alpha \right)$. Here, $J_n(x)$ are Bessel functions of the first kind. 

\clearpage
\bibliography{lib_final}

\end{document}